\documentstyle[12pt]{article}
\begin{document}
\large

\begin{center} {\large\bf ROLE OF NONPERTURBATIVE EFFECTS IN DEEP
INELASTIC SCATTERING REVISITED} \end{center}
\vskip 1em
\begin{center} {\large Felix M. Lev} \end{center}
\vskip 1em
\begin{center} {\it Laboratory of Nuclear
Problems, Joint Institute for Nuclear Research, Dubna, Moscow region
141980 Russia (E-mail:  lev@nusun.jinr.dubna.su)} \end{center}
\vskip 1em
\begin{abstract}
 Restrictions imposed on the (electromagnetic or weak) current operator
by its commutation relations with the representation operators of the
Poincare group are considered in detail. We argue that the present
theory of deep inelastic scattering based on perturbative QCD does not
take into account the dependence of the current operator on the
nonperturbative part of the quark-gluon interaction which cannot be
neglected even in leading order in $1/Q$, where $Q$ is the magnitude of
the momentum transfer.

\begin{flushleft} PACS: 11.40.Dw, 13.60.Hb \end{flushleft}

\begin{flushleft} Key words: deep inelastic scattering, nonperturbative
effects, operator product expansion. \end{flushleft}

\end{abstract}

\section{Introduction}
\label{S1}

 The present theory of deep inelastic scattering (DIS) has proven
rather successful in describing many experimental data. This theory is
based
on two approaches which are the complement of one another. In the first
approach (see e.g. ref. \cite{LP} and references therein) one assumes
that only Feynman diagrams from a certain class dominate in DIS, and in
the second approach DIS is considered in the
framework of the operator product expansion (OPE) \cite{Wil}.
Although the assumptions used in the both approaches are natural, the
problem of their substantiation remains since we do not know how to
work with QCD beyond perturbation theory. In particular, the OPE has
been proved only in perturbation theory \cite{Br} and its validity beyond
that theory is problematic (see the discussion in ref. \cite{Nov} and
references therein).

 At the same time, it is strange that almost all authors
investigating DIS do not pay attention to restrictions
imposed on the (electromagnetic or weak) current operator
by its commutation relations with the representation operators of the
Poincare group. In the present paper we investigate these restrictions
in detail. The paper is organized as follows. In Secs. \ref{S2} and
\ref{S3} we discuss the general properties of the current operator in
quantum field theory and the properties derived in the framework of
canonical formalism. As shown in Sec. \ref{S4}, the latter properties
are not reliable since in some cases they are incompatible with
Lorentz invariance. In Sec. \ref{S5} we apply these results to DIS and
show that the nonperturbative part of the current operator contributes
to deep inelastic scattering even in leading order in $1/Q$ where $Q$ is
the magnitude of the momentum transfer.

\section{Relativistic invariance of the current operator}
\label{S2}

 In any relativistic quantum theory the system under consideration is
described by some (pseudo)unitary representation of the Poincare group.
The current operator ${\hat J}^{\mu}(x)$ for
this system (where $\mu=0,1,2,3$ and $x$ is a point in Minkowski space)
should satisfy the following necessary conditions.

 Let ${\hat U}(a)=exp(\imath {\hat P}_{\mu}a^{\mu})$ be
the representation operator corresponding to the displacement of the origin
in spacetime translation of Minkowski
space by the four-vector $a$. Here ${\hat P}=({\hat P}^0,{\hat {\bf
P}})$ is the operator of the four-momentum, ${\hat P}^0$ is the
Hamiltonian, and ${\hat {\bf P}}$ is the operator of ordinary
momentum. Let also ${\hat U}(l)$ be the representation operator
corresponding to $l\in SL(2,C)$. Then
\begin{equation}
{\hat U}(a)^{-1}{\hat  J}^{\mu}(x){\hat U}(a)=
{\hat J}^{\mu}(x-a)
\label{1}
\end{equation}
\begin{equation}
{\hat U}(l)^{-1}{\hat J}^{\mu}(x){\hat U}(l)=L(l)^{\mu}_{\nu}
{\hat J}^{\nu}(L(l)^{-1}x)
\label{2}
\end{equation}
where $L(l)$ is the element of the Lorentz group corresponding to $l$
and a sum over repeated indices $\mu,\nu=0,1,2,3$ is assumed.

 Let ${\hat M}^{\mu\nu}$ (${\hat M}^{\mu\nu}=-{\hat M}^{\nu\mu}$) be the
representation generators of the Lorentz group.  Then, as follows from
Eq. (\ref{2}), Lorentz invariance of the current operator implies
\begin{equation}
[{\hat M}^{\mu\nu},
{\hat J}^{\rho}(x)]= -\imath\{(x^{\mu}\partial^{\nu}-x^{\nu}\partial^{\mu})
{\hat J}^{\rho}(x)+g^{\mu\rho}{\hat J}^{\nu}(x)-g^{\nu\rho}
{\hat J}^{\mu}(x)\}
\label{3}
\end{equation}
where $g^{\mu\nu}$ is the metric tensor in Minkowski space.

 The operators ${\hat P}^{\mu},{\hat M}^{\mu\nu}$ act in the
scattering space of the system under consideration. In QED the electrons,
positrons and photons are the fundamental particles, and the scattering
space is the space of these almost free particles ("in" or "out" space).
Therefore it is sufficient to deal only with ${\hat P}_{ex}^{\mu},
{\hat M}_{ex}^{\mu\nu}$ where "ex" stands either for "in" or "out".
However in QCD the scattering space by no means can be considered as a
space of almost free fundamental particles --- quarks and gluons. For
example, even if the scattering space consists of one particle (say the
nucleon), this particle is the bound state of quarks and gluons, and the
 operators ${\hat P}^{\mu},{\hat M}^{\mu\nu}$ considerably differ from
the corresponding free operators $P^{\mu},M^{\mu\nu}$. It is well-known
that perturbation theory does not apply to bound states and therefore
${\hat P}^{\mu}$ and ${\hat M}^{\mu\nu}$ cannot be determined in the
framework of perturbation theory. For these reasons we will be interested
in cases when the representation operators in Eqs. (\ref{1}) and (\ref{2})
correspond to the full generators ${\hat P}^{\mu},{\hat M}^{\mu\nu}$.

 Strictly speaking, the notion of current is not necessary if the theory
is complete. For example, in QED there exist unambiguous prescriptions
for calculating the elements of the S-matrix to any desired order of
perturbation theory and this is all we need. It is believed that this
notion is useful for describing the electromagnetic or weak properties
of strongly interacted systems. It is sufficient to know the matrix
elements $\langle \beta|{\hat J}^{\mu}(x)|\alpha \rangle$ of the
operator ${\hat J}^{\mu}(x)$ between the (generalized) eigenstates of the
operator ${\hat P}^{\mu}$ such that ${\hat P}^{\mu}|\alpha\rangle =
P_{\alpha}^{\mu}|\alpha\rangle$, ${\hat P}^{\mu}|\beta\rangle=
P_{\beta}^{\mu}|\beta\rangle$. It is usually assumed that, as a
consequence of Eq. (\ref{1})
\begin{equation}
\langle \beta|{\hat J}^{\mu}(x)|\alpha \rangle=exp[\imath(P_{\beta}^{\nu}-
P_{\alpha}^{\nu})x_{\nu}] \langle \beta|{\hat J}^{\mu}|\alpha \rangle
\label{4}
\end{equation}
where formally ${\hat J}^{\mu}\equiv {\hat J}^{\mu}(0)$. Therefore
in the absence of a complete theory we can consider the less fundamental
problem of investigating the properties of the operator ${\hat J}^{\mu}$.
From the mathematical point of view this implies that we treat
${\hat J}^{\mu}(x)$ not as a four-dimensional operator distribution, but
as a "nonlocal" operator satisfying the condition
\begin{equation}
{\hat J}^{\mu}(x)=exp(\imath {\hat P}x){\hat J}^{\mu}
exp(-\imath {\hat P}x)
\label{5}
\end{equation}

 The standpoint that the current operator should not be treated on the
same footing as the fundamental local fields is advocated by several
authors in their investigations on current algebra (see, for example,
ref. \cite{AFFR}). One of the arguments is that, for example, the
canonical current operator in QED is given by \cite{AB}
\begin{equation}
{\hat J}^{\mu}(x)={\cal N} \{{\hat {\bar \psi}}(x)\gamma^{\mu}{\hat
\psi}(x)\}=\frac{1}{2}[{\hat {\bar \psi}}(x),\gamma^{\mu}{\hat \psi}(x)]
\label{6}
\end{equation}
(where ${\cal N}$ stands for the normal product and ${\hat \psi}(x)$ is
the Heisenberg operator of the Dirac field), but this expression is not a
well-definition of a local operator. Indeed, Eq. (\ref{6}) involves the
product of two local field operators at coinciding points, i.e.
${\hat J}^{\mu}(x)$ is a composite operator.
 The problem of the correct definition of the product of two local
operators at coinciding points is known as
the problem of constructing the composite operators (see e.g. ref.
\cite{Zim}). So far this problem has been solved only in the framework
of perturbation theory for special models. When perturbation theory
does not apply the usual prescriptions are to separate the arguments
of the operators in question and to define the composite operator as
a limit of nonlocal operators when the separation goes to zero (see e.g.
ref. \cite{J} and references therein). Since we do not know how to
work with quantum field theory beyond perturbation theory, we do not
know what is the correct prescription.

 It is well-known (see, for example, ref. \cite{J}) that it is possible
to add to the current operator the term $\partial_{\nu}X^{\mu\nu}(x)$
where $X^{\mu\nu}(x)$ is some operator antisymmetric in $\mu$ and $\nu$.
However it is usually believed \cite{J} that the electromagnetic and
weak current operators of a strongly interacted system are given by the
canonical quark currents the form of which is similar to that in Eq.
(\ref{6}).

 We will not insist on the interpretation of the current operator
according to Eq. (\ref{5}) and will not use this expression in the
derivation of the formulas, but in some cases the notion of
${\hat J}^{\mu}$
makes it possible to explain the essence of the situation clearly.
A useful heuristic expressions which follows from Eqs. (\ref{3}) and
(\ref{5}) is
\begin{equation}
[{\hat M}^{\mu\nu},
{\hat J}^{\rho}]= -\imath (g^{\mu\rho}{\hat J}^{\nu}-g^{\nu\rho}
{\hat J}^{\mu})
\label{7}
\end{equation}

\section{Canonical quantization and the forms of relativistic
dynamics}
\label{S3}

 In the standard formulation of quantum field theory the operators
${\hat P}_{\mu},{\hat M}_{\mu\nu}$ are given by
\begin{equation}
{\hat P}_{\mu}=\int\nolimits {\hat T}_{\mu}^{\nu}(x)d\sigma_{\nu}(x),\quad
{\hat M}_{\mu\nu}=\int\nolimits {\hat M}_{\mu\nu}^{\rho}(x)d\sigma_{\rho}(x)
\label{8}
\end{equation}
where ${\hat T}_{\mu}^{\nu}(x)$ and ${\hat M}_{\mu\nu}^{\rho}(x)$ are the
energy-momentum and angular momentum tensors and
$d\sigma_{\mu}(x)=\lambda_{\mu}\delta(\lambda x-\tau)d^4x$ is
the volume element of the space-like hypersurface defined by the time-like
vector $\lambda \quad (\lambda^2=1)$ and the evolution parameter $\tau$.
In turn, these tensors are fully defined by the classical Lagrangian
and the canonical commutation relations on the hypersurface
$\sigma_{\mu}(x)$. In this connection we note that in canonical
formalism the quantum fields are supposed to be distributions only
relative the three-dimensional variable characterizing the points of
$\sigma_{\mu}(x)$ while the dependence on the variable describing the
distance from $\sigma_{\mu}(x)$ is usual \cite{BLOT}.

 In spinor QED we define $V(x)=-L_{int}(x)=e{\hat J}^{\mu}(x)
{\hat A}_{\mu}(x)$, where $L_{int}(x)$ is the quantum interaction
Lagrangian, $e$ is the (bare) electron charge and ${\hat A}_{\mu}(x)$ is
the operator of the Maxwell field (let us note that if
${\hat J}^{\mu}(x)$ is treated as a composite operator then the product
of the operators entering into $V(x)$ should be correctly defined).

 At this stage it is not necessary to require that ${\hat J}^{\mu}(x)$ is
given by Eq. (\ref{6}), but the key assumption in the canonical
formulation of QED is that ${\hat J}^{\mu}(x)$ is
constructed only from ${\hat \psi}(x)$ (i.e. there is no dependence on
${\hat A}_{\mu}(x)$ and the derivatives of the fields ${\hat A}_{\mu}(x)$
and ${\hat \psi}(x)$). Then the canonical result derived in several
well-known textbooks and monographs (see, for example, ref.
\cite{AB}) is
\begin{equation}
{\hat P}^{\mu}=P^{\mu}+\lambda^{\mu}\int\nolimits V(x)
\delta(\lambda x-\tau)d^4x
\label{9}
\end{equation}
\begin{equation}
{\hat M}^{\mu\nu}=M^{\mu\nu}+\int\nolimits V(x)
(x^{\nu}\lambda^{\mu}-x^{\mu}\lambda^{\nu})
\delta(\lambda x-\tau)d^4x
\label{10}
\end{equation}
 It is important to note that if $A^{\mu}(x)$, $J^{\mu}(x)$ and
$\psi(x)$ are the corresponding free operators then
${\hat A}^{\mu}(x)=A^{\mu}(x)$, ${\hat J}^{\mu}(x)=J^{\mu}(x)$ and
${\hat \psi}(x)=\psi(x)$ if $x\in \sigma_{\mu}(x)$.

  As pointed out by Dirac \cite{Dir}, any physical system can be
described in different forms of relativistic dynamics. By definition,
the description in the point form implies that the operators
${\hat U}(l)$ are the same as for noninteracting particles, i.e.
${\hat U}(l)=U(l)$ and ${\hat M}^{\mu\nu}=M^{\mu\nu}$, and thus
interaction terms can be present only in the four-momentum operators
${\hat P}$ (i.e. in the general case ${\hat P}^{\mu}\neq P^{\mu}$ for
all $\mu$). The description in the instant form implies that the
operators of ordinary momentum and angular momentum do not depend on
interactions, i.e. ${\hat {\bf P}}={\bf P}$, ${\hat {\bf M}}={\bf M}$
$({\hat {\bf M}}=({\hat M}^{23},{\hat M}^{31},{\hat M}^{12}))$, and
therefore interaction terms may be present only in ${\hat P}^0$ and the
generators of the Lorentz boosts ${\hat {\bf N}}=({\hat M}^{01},
{\hat M}^{02},{\hat M}^{03})$. In the front form with the marked $z$
axis we introduce the + and - components of the four-vectors as $x^+=
(x^0+x^z)/\sqrt{2}$, $x^-=(x^0-x^z)/\sqrt{2}$. Then we require that
the operators ${\hat P}^+,{\hat P}^j,{\hat M}^{12},{\hat M}^{+-},
{\hat M}^{+j}$ $(j=1,2)$ are the same as the corresponding free
operators, and therefore interaction terms may be present only in the
operators ${\hat M}^{-j}$ and ${\hat P}^-$.

 In quantum field theory the form of dynamics depends on the choice of
the hypersurface $\sigma_{\mu}(x)$. The representation generators of
the subgroup which leaves this hypersurface invariant are free since
the transformations from this subgroup do not involve dynamics. Therefore
it is reasonable to expect that Eqs. (\ref{9}) and (\ref{10}) give the
most general form of the Poincare group representation generators in
quantum field theory if the fields are quantized on the hypersurface
$\sigma_{\mu}(x)$, but in the general case the relation between $V(x)$
and $L_{int}(x)$ is not so simple as in QED. The fact that the operators
$V(x)$ in Eqs. (\ref{9}) and (\ref{10}) are the same follows from the
commutation relations between ${\hat P}^{\mu}$ and ${\hat M}^{\mu\nu}$.

 The most often considered case is $\tau =0$, $\lambda =(1,0,0,0)$. Then
$\delta(\lambda x-\tau)d^4x=d^3{\bf x}$ and the integration in Eqs.
(\ref{9}) and (\ref{10}) is taken over the hyperplane $x^0=0$. Therefore,
as follows from these expressions, ${\hat {\bf P}}={\bf P}$ and
${\hat {\bf M}}={\bf M}$. Hence such a choice of $\sigma_{\mu}(x)$
leads to the instant form \cite{Dir}.

  The front form can be formally obtained from Eqs. (\ref{9}) and
(\ref{10}) as follows. Consider the vector $\lambda$ with the components
\begin{equation}
\lambda^0=\frac{1}{(1-v^2)^{1/2}},\quad \lambda^j=0,\quad
\lambda^3=-\frac{v}{(1-v^2)^{1/2}}\quad (j=1,2)
\label{11}
\end{equation}
Then taking the limit $v\rightarrow 1$ in Eqs. (\ref{9}) and
(\ref{10}) we get
\begin{eqnarray}
&&{\hat P}^{\mu}=P^{\mu}+\omega^{\mu}\int\nolimits V(x)
\delta(x^+)d^4x,\nonumber\\
&&{\hat M}^{\mu\nu}=M^{\mu\nu}+\int\nolimits V(x)
(x^{\nu}\omega^{\mu}-x^{\mu}\omega^{\nu})
\delta(x^+)d^4x
\label{12}
\end{eqnarray}
where the vector $\omega$ has the components $\omega^-=1$,
$\omega^+=\omega^j=0$. It is obvious that the generators (\ref{12}) are
given in the front form and that's why Dirac \cite{Dir} related this form
to the choice of the light cone $x^+=0$.

 In ref. \cite{Dir} the point form was related to the hypersurface
$t^2-{\bf x}^2>0,\,t>0$, but as argued by Sokolov \cite{Sok},
the point form should be related to the hyperplane orthogonal to the
four-velocity of the
system under consideration. We shall not discuss this question in the
present paper.

\begin{sloppypar}
\section{Incompatibility of canonical formalism with Lorentz invariance
for spinor fields}
\label{S4}
\end{sloppypar}

 In canonical formalism the key property of the current operator for
the spinor field is that ${\hat J}^{\mu}(x)=J^{\mu}(x)$ if
$x\in \sigma_{\mu}(x)$. The purpose of this section is to show that
this property is not correct since it is incompatible with Lorentz
invariance.

 A possible objection against the derivation of Eqs. (\ref{9}) and
(\ref{10}) is that the product of local operators at
one and the same value of $x$ is not a well-defined object. For
example, if $x^0=0$ then following Schwinger \cite{Schw}, instead of
Eq. (\ref{6}), one can define $J^{\mu}({\bf x})$ as the limit of the
operator
\begin{equation}
J^{\mu}({\bf x})=\frac{1}{2}[{\bar \psi}({\bf x}+\frac{{\bf l}}{2}),
\gamma^{\mu}exp(\imath e
\int_{{\bf x}-\frac{{\bf l}}{2}}^{{\bf x}+\frac{{\bf l}}{2}}
{\bf A}({\bf x}')d{\bf x}')
{\bar \psi}({\bf x}-\frac{{\bf l}}{2})]
\label{13}
\end{equation}
when ${\bf l}\rightarrow 0$, the limit should be taken only at the
final stage of calculations and in the general case the time
components of the arguments of ${\hat {\bar \psi}}$ and ${\hat \psi}$
also differ each other (the contour integral in this expression is
needed to conserve gauge invariance). Therefore there is a "hidden"
dependence of ${\hat J}^{\mu}(x)$ on ${\hat A}^{\mu}(x)$ and hence Eqs.
(\ref{9}) and (\ref{10}) are incorrect.

 However, any attempt to separate the arguments of the ${\hat \psi}$
operators in ${\hat J}^{\mu}(x)$ immediately results in breaking of
locality. In particular, at any ${\bf l}\neq 0$ in Eq. (\ref{13}) the
Lagrangian is nonlocal. We do not think that locality is a primary
physical condition, but once the Lagrangian is nonlocal, the whole
edifice of local quantum field theory (including canonical
formalism) becomes useless. Meanwhile the only known way of constructing
the generators ${\hat P}^{\mu},{\hat M}^{\mu\nu}$ in local quantum
field theory is canonical formalism. For these reason we first consider
the results which formally follow from canonical formalism and show
that these results are inconsistent.

\begin{sloppypar}
 In addition to the properties discussed above, the current operator
should also satisfy the continuity equation $\partial
{\hat J}^{\mu}(x)/\partial x^{\mu}=0$. As follows from this equation and
Eq. (\ref{1}), $[{\hat  J}^{\mu}(x),{\hat P}_{\mu}]=0$.
The canonical formalism in the instant form implies that if $x^0=0$ then
${\hat  J}^{\mu}({\bf x})=J^{\mu}({\bf x})$. Since $J^{\mu}({\bf x})$
satisfies the condition $[J^{\mu}({\bf x}),P_{\mu}]=0$, it follows from
Eq. (\ref{9}) that if ${\hat P}^{\mu}=P^{\mu}+V^{\mu}$ then the
continuity equation is satisfied only if
\begin{equation}
[V^0,J^0({\bf x})]=0
\label{14}
\end{equation}
where
\begin{equation}
V^0=\int\nolimits V({\bf x})d^3{\bf x},\quad V({\bf x}) =-e
{\bf A}({\bf x}){\bf J}({\bf x})
\label{15}
\end{equation}
We take into account the fact that the canonical quantization on the
hypersurface $x^0=0$ implies that $A^0({\bf x})=0$.
\end{sloppypar}

 As follows from Eqs. (\ref{1}) and (\ref{3}), the commutation relation
between the operators ${\hat M}^{0i}$ $(i=1,2,3)$ and $J^0({\bf x})$
should have the form
\begin{equation}
[{\hat M}^{0i},J^0({\bf x})]=-x^i[{\hat P}^0,J^0({\bf x})]-
\imath J^i({\bf x})
\label{16}
\end{equation}
Since
\begin{equation}
[M^{0i},J^0({\bf x})]=-x^i[P^0,J^0({\bf x})]-\imath J^i({\bf x})
\label{17}
\end{equation}
it follows from Eqs. (\ref{10}), (\ref{14}) and (\ref{15}) that Eq.
(\ref{16}) is satisfied if
\begin{equation}
\int\nolimits y^i {\bf A}({\bf y})[{\bf J}({\bf y}),J^0({\bf x})]
d^3{\bf y}=0
\label{18}
\end{equation}
It is well-known that if the standard equal-time commutation relations
are used naively then the commutator in Eq. ({\ref{18}) vanishes and
therefore this equation is satisfied. However when ${\bf x}\rightarrow
{\bf y}$ this commutator involves the product of four Dirac fields at
${\bf x}={\bf y}$. The famous Schwinger result \cite{Schw} is that if
the current operators in question are given by Eq. (\ref{13}) then
\begin{equation}
[J^i({\bf y}),J^0({\bf x})]=C\frac{\partial}{\partial x^i}
\delta({\bf x}-{\bf y})
\label{19}
\end{equation}
where $C$ is some (infinite) constant. Therefore Eq. ({\ref{18}) is not
satisfied and the current operator ${\hat J}^{\mu}(x)$ constructed in
the framework of canonical formalism does not satisfy Lorentz invariance.

 At the same time, Eq. (\ref{19}) is compatible with Eqs. (\ref{14}) and
(\ref{15}) since $div({\bf A}({\bf x}))=0$. One can also expect that the
commutator $[{\hat M}^{0i},J^k({\bf x})]$ is compatible with Eq. (\ref{3}).
This follows from the fact \cite{GJ} that if Eq. (\ref{19}) is
satisfied then the commutator $[J^i({\bf x}),J^k({\bf y})]$ does not
contain derivatives of the delta function.

 While the arguments given in
Ref. \cite{Schw} prove that the commutator in Eq. (\ref{19}) cannot vanish,
one might doubt whether the singularity of the commutator is indeed given
by the right hand side of this expression. Of course, at present any
method of calculating such a commutator is model dependent, but
the result that canonical formalism  is incompatible with Lorentz
invariance (see Eq. (\ref{16}))
follows in fact only from algebraic considerations. Indeed, Eqs.
(\ref{14}), (\ref{16}) and (\ref{17}) imply that if ${\hat M}^{\mu\nu}=
M^{\mu\nu}+V^{\mu\nu}$ then
\begin{equation}
[V^{0i},J^0({\bf x})]=0
\label{20}
\end{equation}

 Since $V^{0i}$ in the instant form is a nontrivial interaction
dependent operator, there is no reason to expect that it commutes with
the free operator $J^0({\bf x})$. Moreover for the analogous reason
Eq. (\ref{14}) will not be satisfied in the general case.

 To better understand the situation in spinor QED it is useful to
consider scalar QED \cite{J2}. The formulation of this theory can
be found, for example, in ref. \cite{IZ}. In contrast with spinor
QED, the Schwinger term in scalar QED emerges canonically \cite{Schw,J}.
We use $\varphi({\bf x})$ to denote the operator of the scalar complex
field at $x^0=0$. The canonical calculation yields
\begin{eqnarray}
&&{\hat J}^0({\bf x})=J^0({\bf x})=\imath[\varphi^*({\bf x}) \pi^*({\bf x})-
\pi({\bf x})\varphi({\bf x})], \nonumber\\
&& {\hat J}^i({\bf x})=J^i({\bf x})-
2eA^i({\bf x})\varphi^*({\bf x}) \varphi({\bf x}),\nonumber\\
&& J^i({\bf x})=\imath [\varphi^*({\bf x})\cdot\partial^i
\varphi({\bf x})-\partial^i \varphi^*({\bf x})\cdot
\varphi({\bf x})]
\label{21}
\end{eqnarray}
where $\pi({\bf x})$ and $\pi^*({\bf x})$ are the operators canonically
conjugated with $\varphi({\bf x})$ and $\varphi^*({\bf x})$ respectively.
In contrast with Eq. (\ref{15}), the operator $V({\bf x})$ in scalar
QED is given by
\begin{equation}
V({\bf x}) =-e {\bf A}({\bf x}){\bf J}({\bf x})
+e^2{\bf A}({\bf x})^2\varphi^*({\bf x})\varphi({\bf x})
\label{22}
\end{equation}
However the last term in this expression does not contribute to the
commutator (\ref{16}). It is easy to demonstrate that as pointed out
in Ref. \cite{J2}, the commutation relations (\ref{3}) in scalar QED
are satisfied in the framework of the canonical formalism.
Therefore the naive treatment of the product of local operators at
coinciding points in this theory is not in conflict with the
canonical commutation relations.
The key difference between spinor QED and scalar QED is that in contrast
with spinor QED, the spatial component of the canonical current operator
is not free if $x^0=0$ (see Eq. (\ref{21})). Just for this
reason the commutator $[{\hat M}^{0i},J^0({\bf x})]$ in scalar QED
agrees with Eq. ({\ref{3}) since the Schwinger term in this commutator
gives the interaction term in ${\hat J}^i({\bf x})$.

 Now let us return to spinor QED. As noted above, the canonical
formalism cannot be used if the current operator is considered as a
limit of the expression similar to that in Eq. (\ref{13}). In
addition, the problem exists what is the correct definition of
$V({\bf x})$ as a composite operator. One might expect that the
correct definition of $J^{\mu}({\bf x})$ and $V({\bf x})$ will result
in appearance of some additional terms in $V({\bf x})$ (and hence in $V^0$
and $V^{0i}$). However it is unlikely that this is the main reason of
the violation of Lorentz invariance. Indeed, as noted above, for only
algebraic reasons it is unlikely that both conditions (\ref{14}) and
(\ref{20}) can be simultaneously satisfied. Therefore, taking into
account the situation in scalar QED, {\it it is natural to conclude that
the main reason of the failure of canonical formalism is that either the
limit of ${\hat J}^{\mu}(x^0,{\bf x})$ when $x^0\rightarrow 0$ does not
exist or this limit is not equal to $J^{\mu}({\bf x})$} (i.e. the
relation ${\hat J}^{\mu}({\bf x})=J^{\mu}({\bf x})$ is incorrect).

 The fact that the relation ${\hat J}^{\mu}({\bf x})=J^{\mu}({\bf x})$
cannot be correct follows from simpler considerations. Indeed, assume
first that this relation is valid. Then we can use canonical formalism
in the framework of which the generator of the gauge transformations is
$div {\bf E}({\bf y}) - J^0({\bf y})$, and if ${\bf J}({\bf x})$ is gauge
invariant then $[div {\bf E}({\bf y}) - J^0({\bf y}),
{\bf J}({\bf x})]=0$. The commutator $[J^0({\bf y}),{\bf J}({\bf x})]$
cannot be equal to zero \cite{Schw} and therefore ${\bf J}({\bf x})$
does not commute with $div {\bf E}({\bf y})$ while the free operator
${\bf J}({\bf x})$ commutes with $div {\bf E}({\bf y})$.
The relation ${\hat J}^{\mu}({\bf x})=J^{\mu}({\bf x})$ also does not
take place in explicitly solvable two-dimensional models \cite{BLOT}.
In addition, once we assume that the field operators on
the hypersurface $\sigma_{\mu}(x)$ are free we immediately are in
conflict with the Haag theorem \cite{Haag,BLOT}. However for our
analysis of the current operator in DIS in Sec. \ref{S5} it is
important that ${\hat J}^{\mu}({\bf x})\neq J^{\mu}({\bf x})$ as a
consequence of Lorentz invariance.

\begin{sloppypar}
By analogy with ref. \cite{Schw} it is easy to show that if $x^+=0$
then the canonical current operator in the front form $J^+(x^-,
{\bf x}_{\bot})$ (we use the subscript $\bot$ to denote the projection
of the three-dimensional vector onto the plane 12) cannot commute with
all the operators $J^i(x^-,{\bf x}_{\bot})$ $(i=-,1,2)$. As easily
follows from the continuity equation and Lorentz invariance (\ref{3}),
the canonical operator $J^+(x^-,{\bf x}_{\bot})$ should satisfy the
relations
\begin{equation}
[V^-,J^+(x^-,{\bf x}_{\bot})]=[V^{-j},J^+(x^-,{\bf x}_{\bot})]=0
\quad (j=1,2)
\label{23}
\end{equation}
By analogy with the above consideration we conclude that these
relations cannot be simultaneously satisfied and therefore
{\it either the limit of ${\hat J}^{\mu}(x^+,x^-,{\bf x}_{\bot})$ when
$x^+\rightarrow 0$ does not exist or this limit is not equal to
$J^{\mu}(x^-,{\bf x}_{\bot})$}. Therefore the canonical light cone
quantization does not render a Lorentz invariant current operator for
spinor fields.
\end{sloppypar}

 Let us also note that if the theory should be invariant under the space
reflection or time reversal, the corresponding representation operators
in the front form ${\hat U}_P$ and ${\hat U}_T$ are necessarily
interaction dependent (this is clear, for example, from the relations
${\hat U}_PP^+{\hat U}_P^{-1}$ = ${\hat U}_TP^+{\hat U}_T^{-1}$ =
${\hat P}^-$). In terms of the operator ${\hat J}^{\mu}$ one can say
that this operator should satisfy the conditions
\begin{equation}
{\hat U}_P({\hat J}^0,{\hat {\bf J}}){\hat U}_P^{-1}=
{\hat U}_T({\hat J}^0,{\hat {\bf J}}){\hat U}_T^{-1}=
({\hat J}^0,-{\hat {\bf J}})
\label{24}
\end{equation}
Therefore it is not clear whether these conditions are compatible with
the relation ${\hat J}^{\mu}=J^{\mu}$. However this difficulty is
a consequence of the difficulty with Eq. (\ref{2}) since, as noted by
Coester \cite{Coes}, the interaction dependence of the operators
${\hat U}_P$ and ${\hat U}_T$ in the front form does not mean that there
are discrete dynamical symmetries in addition to the rotations about
transverse axes.
Indeed, the discrete transformation $P_2$ such that
$P_2\, x:= \{x^0,x_1,-x_2,x_3\}$ leaves the light front $x^+=0$ invariant.
The full space reflection $P$ is the product of $P_2$ and a rotation about
the 2-axis by $\pi$. Thus it is not an independent dynamical transformation
in addition to the rotations about transverse axes.
Similarly the transformation $TP$ leaves $x^+=0$ invariant and
$T=(TP)P_2R_2(\pi)$.

 In terms of the operator ${\hat J}^{\mu}$ the results of this section
are obvious. Indeed, since at $x=0$ the Heisenberg and Schrodinger
pictures coincide then in view of Eq. (\ref{6}) one might think that
the operator ${\hat J}^{\mu}$ is free, i.e. ${\hat J}^{\mu}=J^{\mu}$.
However there is no reason for the interaction terms in $M^{\mu\nu}$
to commute with the free operator $J^{\mu}$ (see Eq. (\ref{7})).
Therefore the results of this section show that the algebraic reasons
based on Eq. (\ref{7}) are more solid than the reasons based on
formal manipulations with local operators, and in the instant and front
forms ${\hat J}^{\mu}\neq J^{\mu}$. Note also that although the model
considered in this section is spinor QED, the above results are not
very important for QED itself since, as pointed out in
Sec. \ref{S2}, in QED it is sufficient to consider only commutators
involving ${\hat P}_{ex}^{\mu}$ and ${\hat M}_{ex}^{\mu\nu}$. However
it will be clear in the next section that the above considerations are
important for investigating the properties of the current operator for
strongly interacting particles.

\section{Current operator in DIS}
\label{S5}

 If $q$ is the momentum transfer in DIS then the DIS cross-section
is fully defined by the hadronic tensor
\begin{equation}
W^{\mu\nu}=\frac{1}{4\pi}\int\nolimits e^{\imath qx} \langle P'|
[{\hat J}^{\mu}(\frac{x}{2}),{\hat J}^{\nu}(-\frac{x}{2})]
|P'\rangle d^4x
\label{25}
\end{equation}
where the initial nucleon state $|P'\rangle$ is the eigenstate of the
operator ${\hat P}$ with the eigenvalue $P'$ and the eigenstate of the
spin operators ${\hat {\bf S}}^2$ and ${\hat S}^z$ which are constructed
from ${\hat M}^{\mu\nu}$. In particular,
${\hat P}^2|P'\rangle =m^2 |P'\rangle$ where $m$ is the nucleon mass.
Therefore the four-momentum operator indeed necessarily depends on the
nonperturbative part of the interaction which is responsible for binding
of quarks and gluons in the nucleon.

  Suppose that the Hamiltonian ${\hat P}^0$ contains the nonperturbative
part of the quark-gluon interaction and consider the well-known relation
$[{\hat M}^{0i},{\hat P}^k]=-\imath \delta_{ik}{\hat P}^0$ $(i,k=1,2,3)$.
Then it is obvious that if ${\hat P}^k=P^k$ then all the operators
${\hat M}^{0i}$ inevitably contain the nonperturbative part and
{\it vice versa}, if ${\hat M}^{0i}=M^{0i}$ then all the operators
${\hat P}^k$ inevitably contain this part.
Therefore in the instant form all the operators ${\hat M}^{0i}$
inevitably depend on the nonperturbative part of the quark-gluon
interaction and in the point form all the operators ${\hat P}^k$
inevitably depend on this part. In the front form the fact that all the
operators ${\hat M}^{-j}$  inevitably depend on the
nonperturbative part follows from the relation
$[{\hat M}^{-j},{\hat P}^l]=-\imath \delta_{jl}{\hat P}^-$ $(j,l=1,2)$.
Of course, the physical results should not depend on the choice of the
form of dynamics and in the general case all ten generators can
depend on the nonperturbative part of the quark-gluon interaction.

 In turn, as follows from Eq. (\ref{3}) and the results of Sec. \ref{S4},
the operators ${\hat J}^{\mu}({\bf x})$ in the instant form and
${\hat J}^{\mu}(x^-,{\bf x}_{\bot})$ in the front one
inevitably depend on the nonperturbative part of the quark-gluon
interaction. If it is possible to define ${\hat J}^{\mu}$ in the
point form then as follows from Eq. (\ref{7}), the relation
${\hat J}^{\mu}=J^{\mu}$ does not contradict Lorentz invariance
but, as follows from Eq. (\ref{5}), the operator ${\hat J}^{\mu}(x)$ in
that form inevitably depend on the nonperturbative part.
The fact that the same operators $({\hat P}^{\mu},{\hat M}^{\mu\nu})$
describe the transformations of both the operator ${\hat J}^{\mu}(x)$
and the state $|P'\rangle$ guaranties that $W^{\mu\nu}$ has the
correct transformation properties.

 We see that the relation between the current operator and the state of
the initial nucleon is highly nontrivial. Meanwhile in the present theory
they are considered separately. In the framework of the approach based
on Feynman diagrams the possibility of the separate consideration
follows from the factorization theorem \cite{ER} which asserts in
particular that the amplitude of the lepton-parton interaction
entering into diagrams dominating in DIS depend only on the hard part
of this interaction. Moreover, in leading order in $1/Q$,
where $Q=|q^2|^{1/2}$, one obtains the parton model up
to anomalous dimensions and perturbative QCD corrections which depend on
$\alpha_s(Q^2)$ where $\alpha_s$ is the QCD running coupling constant.

 It is well-known that the parton model is equivalent to impulse
approximation (IA) in the infinite momentum frame (IMF). This fact is in
agreement with our experience in conventional nuclear and atomic physics
according to which in processes with high momentum transfer the effect of
binding is not important and the current operator can be taken in IA.
However this experience is based on the nonrelativistic
quantum mechanics where only the Hamiltonian is interaction dependent
and the other nine generators of the Galilei group are free. Note
also that in the nonrelativistic case the kinetic energies and the
interaction operators in question are much smaller than the masses of
the constituents.

 Let us now discuss the following question. Since the current
operator depends on the nonperturbative part of the quark-gluon
interaction then this operator depends on the integrals from the quark
and gluon field operators over the region of large distances where the
QCD running coupling constant $\alpha_s$ is large. Is this property
compatible with locality? In the framework of canonical formalism
compatibility is obvious but, as shown in the preceding section,
the results based on canonical formalism are not reliable.
Therefore it is not clear whether in QCD it is possible to construct
local electromagnetic and weak current operators beyond perturbation
theory. However the usual motivation of the parton model is that,
as a consequence of asymptotic freedom (i.e. the fact
that $\alpha_s(Q^2)\rightarrow 0$ when $Q^2 \rightarrow \infty$),
the partons in the IMF are almost free and therefore, at least in
leading order in $1/Q$, the nonperturbative part of ${\hat J}^{\mu}(x)$
is not important. We will now consider whether this property can be
substantiated in the framework of the OPE.

 In this framework the commutator of the currents
entering into Eq. (\ref{25}) can be written symbolically as
\begin{equation}
[{\hat J}(\frac{x}{2}),{\hat J}(-\frac{x}{2})]
= \sum_{i} C_i(x^2) x_{\mu_1}\cdots x_{\mu_n}
{\hat O}_i^{\mu_1\cdots \mu_n}
\label{26}
\end{equation}
where $C_i(x^2)$ are the $c$-number Wilson coefficients while the
operators ${\hat O}_i^{\mu_1\cdots \mu_n}$ depend only on field
operators and their covariant derivatives at the origin of Minkowski
space and have the same form as in perturbation theory. The basis for
twist two operators contains in particular
\begin{equation}
{\hat O}_V^{\mu}={\cal N} \{{\hat {\bar \psi}}(0)\gamma^{\mu}{\hat
\psi}(0)\} \quad
{\hat O}_A^{\mu}={\cal N} \{{\hat {\bar \psi}}(0)\gamma^{\mu}
\gamma^5{\hat \psi}(0)\}
\label{27}
\end{equation}
where for simplicity we do not write flavor operators and color and
flavor indices.

 As noted above, the operator ${\hat J}^{\mu}(x)$ necessarily depends
on the nonperturbative part of the quark-gluon interaction  while Eq.
(\ref{26}) has been proved only in the framework of
perturbation theory. Therefore if we use Eq. (\ref{26}) in DIS we have to
assume that either nonperturbative effects are not important to some
orders in $1/Q$ and then we can use Eq. (\ref{26}) only
to these orders (see e.g. ref. \cite{Jaffe}) or it is possible to use Eq.
(\ref{26}) beyond perturbation theory. The question also arises whether
Eq. (\ref{26}) is valid in all the forms of dynamics (as it should be if
it is an exact operator equality) or only in some forms.

 In the point form all the components of ${\hat P}$ depend on the
nonperturbative part of the quark-gluon interaction
and therefore, in view of Eqs. (\ref{1}) or (\ref{5}), it is not clear
why there is no nonperturbative part in the $x$ dependence of the right
hand side of Eq. (\ref{26}), or if (for some reasons) it is possible to 
include the nonperturbative part only into the operators ${\hat O}_i$ 
then why they have the same form as in perturbation theory.

 One might think that in the front form the $C_i(x^2)$ will be the
same as in perturbation theory due to the following reasons. The
value of $q^-$ in DIS is very large and therefore only a small
vicinity of the light cone $x^+=0$ contributes to the integral (\ref{25}).
The only dynamical component of ${\hat P}$ is ${\hat P}^-$ which enters
into Eq. (\ref{26}) only in the combination ${\hat P}^-x^+$. Therefore
the dependence of ${\hat P}^-$ on the nonperturbative part of the
quark-gluon interaction is of no importance.  These considerations
are not convincing since the integrand is a singular function and the
operator ${\hat J}^{\mu}(x^-,{\bf x}_{\bot})$ in the front form depends
on the nonperturbative part, but nevertheless we assume that Eq.
(\ref{26}) in the front form is valid.

 If we assume as usual that there is no problem with the convergence
of the OPE series then experiment makes it possible to measure
each matrix element $\langle P'|{\hat O}_i^{\mu_1\cdots \mu_n}|P'\rangle$.
Let us consider, for example, the matrix element
$\langle P'|{\hat O}_V^{\mu}|P'\rangle$. It transforms as a four-vector
if the Lorentz transformations of ${\hat O}_V^{\mu}$ are described by the
operators ${\hat M}^{\mu\nu}$ describing the transformations of
$|P'\rangle$, or in other words, by analogy with Eq. (\ref{7})
\begin{equation}
[{\hat M}^{\mu\nu},{\hat O}_V^{\rho}]=-\imath (g^{\mu\rho}{\hat O}_V^{\nu}-
g^{\nu\rho} {\hat O}_V^{\mu})
\label{28}
\end{equation}
It is also clear that Eq. (\ref{28}) follows from Eqs. (\ref{1}),
(\ref{3}) and (\ref{25}).
Since the ${\hat M}^{-j}$ in the front form depend on the
nonperturbative part of the quark-gluon interaction, the results of
Sec. \ref{S4} make it possible to conclude that at least
some components ${\hat O}_V^{\mu}$, and analogously some components
${\hat O}_i^{\mu_1\cdot \mu_n}$, also depend on the
nonperturbative part. Since Eq. (\ref{28}) does not contain any $x$ or
$q$ dependence, this conclusion has nothing to do with asymptotic
freedom and is valid even in leading order in $1/Q$ (in contrast with the
statement of the factorization theorem \cite{ER}).
Since the struck quark is not free but interacts nonperturbatively with
the rest of the target then, in terminology of ref. \cite{LP}, not
only "handbag" diagrams dominate in DIS but some of "cat ears" diagrams
or their sums are also important (in other words, even the notion of
struck quark is questionable).

 Since the operators ${\hat O}_i^{\mu_1...\mu_n}$ depend on the
nonperturbative part of the quark-gluon interaction, then by analogy
with the above considerations we conclude that the operators in Eq.
(\ref{27}) are ill-defined and the correct expressions for them involve
integrals from the field operators over large distances where the QCD
coupling constant is large. Therefore it is not clear whether
the operators ${\hat O}_i^{\mu_1...\mu_n}$ are local and whether the
Taylor expansion at $x=0$ is correct, but even it is, the expressions for
${\hat O}_i^{\mu_1...\mu_n}$ will depend on higher twist operators which
contribute even in leading order in $1/Q$.

\section{Discussion}
\label{S6}

 As follows from the results of Sec. \ref{S4},
the current operator nontrivially depends on the nonperturbative part
of the interaction responsible for binding of quarks and gluons in the
nucleon. Then the problem arises whether it is possible to construct
a local current operator ${\hat J}^{\mu}(x)$ beyond perturbation
theory and whether the nonperturbative part of the interaction
entering into ${\hat J}^{\mu}(x)$ contributes to DIS.
Our consideration shows that the dependence of ${\hat J}^{\mu}(x)$ on the
nonperturbative part of the interaction makes the OPE problematic.
Nevertheless we assume that Eq. (\ref{26}) is valid beyond perturbation
theory but no form of the operators ${\hat O}_i^{\mu_1...\mu_n}$ is
prescribed. Then we come to conclusion that the nonperturbative part
contributes to DIS even in leading order in $1/Q$.

 To understand whether the OPE is valid beyond perturbation theory
several authors (see e.g. ref. \cite{Nov} and references therein)
investigated some two-dimensional models and came to different
conclusions. We will not discuss the arguments of these authors but
note that the Lie algebra of the Poincare group for 1+1 space-time
is much simpler than for 3+1 one. In particular, the Lorentz group is
one-dimensional and in the front form the operator $M^{+-}$ is free.
Therefore Eqs. (\ref{7}) and (\ref{28}) in the "1+1 front form" do not
make it possible to conclude that the operators ${\hat J}^{\mu}$ and
${\hat O}_V^{\mu}$ necessarily depend on the nonperturbative part of
the quark-gluon interaction. At the same time the full space
reflection $P$ in the 1+1 front form is an independent dynamical
transformation, in contrast with the situation in the 3+1 front form
(see Sec. \ref{S4}).

 Note also that in solvable models considered in the literature the
operators ${\hat P}^{\mu},{\hat M}^{\mu\nu}$ were not explicitly
constructed; in particular, it is not clear whether the current
operators in these models satisfy the conditions (\ref{1},\ref{2}).

 Since the operators ${\hat O}_i^{\mu_1...\mu_n}$ in Eq. (\ref{27})
should depend
on the nonperturbative part of the quark-gluon interaction then,
as noted above, there is no reason to think that these operators
are local but even if they are then twist (dimension minus spin) no
longer determines in which order in $1/Q$ the corresponding
operator contributes to DIS. This is clear from the fact that the
dependence on the nonperturbative part implies that we have an
additional parameter $\Lambda$ with the dimension of momentum
where $\Lambda$ is the characteristic momentum at which
$\alpha_s(\Lambda^2)$ is large.

 Nevertheless if we assume that (for some reasons) Eq. (\ref{26}) is
still
valid and consider only the $q^2$ evolution of the structure functions
then all the standard results remain. Indeed the only information
about the operators ${\hat O}_i^{\mu_1...\mu_n}$ we need is their tensor
structure since we should correctly parametrize the matrix elements
$\langle P'|{\hat O}_i^{\mu_1\cdots \mu_n}|P'\rangle$. However the
derivation of sum rules in DIS requires additional assumptions.

 Let us consider sum rules in DIS in more details. It is well-known
that they are derived with different extent of rigor. For example,
the Gottfried and Ellis-Jaffe sum rules \cite{Got} are essentially
based on model assumptions, the sum rule \cite{Adl} was originally
derived in the framework of current algebra for the time component
of the current operator while the sum rules \cite{Bjor} also involve
the space components. As shown in Sec. \ref{S4}, the operator
${\hat {\bf J}}({\bf x})$ is necessarily interaction dependent
and there exist models in which ${\hat J}^0({\bf x})$ is free.
Therefore in the
framework of current algebra the sum rule \cite{Adl} is substantiated
in greater extent than the sum rules \cite{Bjor} (for a detailed
discussion see refs. \cite{GM,J}). Now the sum rules \cite{Adl,Bjor}
are usually considered in the framework of the OPE and they
have the status of fundamental relations which in fact unambiguously
follow from QCD. However the important assumption in deriving the sum
rules is that the expression for ${\hat O}_V^{\mu}$ coincides with
${\hat J}^{\mu}$, the expression for ${\hat O}_A^{\mu}$ coincides with
the axial current operator ${\hat J}_A^{\mu}$ etc.
(see Eqs. (\ref{6}) and (\ref{27})). Our results
show that this assumption has no physical ground. Therefore although
(for some reasons) there may exist sum rules which are satisfied with
a good accuracy, the statement that the sum rules \cite{Adl,Bjor}
unambiguously follow from QCD is not substantiated.

 For comparing the theoretical predictions for the sum rules with
experimental data it is also very important to calculate effects in
next-to-leading order in $1/Q$. As shown in ref. \cite{Mart} there
exist serious difficulties in calculating such effects in the
framework of the OPE, and the authors of ref. \cite{Mart} are very
pessimistic about the possibility to overcome these difficulties
(while in our approach problems exist even in the leading order).

 The current operator satisfying Eqs. (\ref{1}) and (\ref{3}) can be
explicitly
constructed for systems with a fixed number of interacting
relativistic particles \cite{lev}. In such models it is clear when
the corresponding results and the results in IA are similar and when
they considerably differ \cite{hep1}.

 We conclude that the present theory of DIS based on perturbative
QCD does not take into account the dependence of the current operator
on the nonperturbative part of the quark-gluon interaction which
cannot be neglected even in leading order in $1/Q$. On the other
hand, as already noted, the present theory has proven rather successful 
in describing many experimental data. It is very important to understand 
why this situation takes place.

\begin{center} {\bf Acknowledgments} \end{center}
\begin{sloppypar}

 The author is grateful to B.L.G.Bakker, F.Coester, R. Van Dantzig,
L.A.Kondratyuk, B.Z.Kopeliovich, M.P.Locher,
S.Mikhailov, P.Muelders, N.N.Nikolaev, E.Pace, R.Petronzio,
R.Rosenfelder, G.Salme, O.Yu.Shevchenko, I.L.Solovtsov and H.J.Weber
for valuable discussions. This work was supported
by grant No. 96-02-16126a from the Russian Foundation for Fundamental
Research.

\end{sloppypar}

\end{document}